\documentclass[10pt,conference]{IEEEtran}

\usepackage[english]{babel} 
\usepackage{amsmath}                
\usepackage{amsfonts}               
\usepackage{amssymb}                
\usepackage{amsopn}                 
\allowdisplaybreaks[3] 
\usepackage{bbm}                    
\usepackage{mathrsfs}               
\usepackage{calc}                   
\usepackage[dvips]{graphicx}        
\usepackage{epsfig}                
\usepackage{psfrag}                 
\usepackage[dvips]{color}           
\usepackage{fancyhdr}               
\usepackage{verbatim}               
\usepackage{exscale}                
\usepackage{research5}
\newtheorem{theorem}{Theorem}

\newcommand{\Cfb}{\dot{C}_{\textnormal{FB}}(0)}

\title{A Channel that Heats Up}

\author{\authorblockN{Tobias Koch ~~~~~ Amos Lapidoth}
\authorblockA{ETH
  Zurich, Switzerland\\
Email: \{tkoch, lapidoth\}@isi.ee.ethz.ch}
\and 
\authorblockN{Paul P.~Sotiriadis}
\authorblockA{Johns Hopkins University, Baltimore, USA\\Email: pps@jhu.edu}}

\begin{document}

\maketitle

\begin{abstract}
  Motivated by on-chip communication, a channel model is proposed where the variance of the additive noise
  depends on the weighted sum of the past channel input powers. For
  this channel, an expression for the capacity per unit cost is derived, and it is shown that the expression holds
  also in the presence of feedback.
\end{abstract}

\section{Introduction}
\label{sec:intro}
Continuous advancement in VLSI technologies has resulted in extremely small
transistor sizes and highly complex microprocessors. However, on-chip
interconnects responsible for on-chip communication have been improved only
moderately. This leads to the ``paradox'' that local information processing is
done very efficiently, but communicating information between on-chip units
is a major challenge.

This work focuses on an emergent issue expected to challenge circuit
development in future technologies. Information communication and processing
is associated with energy dissipation into heat which raises the
temperature of the transmitter/receiver or processing
devices; moreover, the
intrinsic device noise level depends strongly and increasingly on
the temperature. Therefore, the total physical structure can be modeled as a
communication channel whose noise level is data dependent. We describe
this mathematically in the following subsection.

\subsection{Channel Model}
\label{sub:channelmodel}
\begin{figure}[t!]
 \centering
 \psfrag{T}[cc][cc]{Transmitter}
 \psfrag{C}[cc][cc]{Channel}
 \psfrag{R}[cc][cc]{Receiver}
 \psfrag{D}[cc][cc]{Delay}
 \psfrag{M}[b][b]{$M$}
 \psfrag{Mh}[b][b]{$\hat{M}$}
 \psfrag{X}[b][b]{$X_k$}
 \psfrag{Y}[b][b]{$Y_k$}
 \psfrag{Yr}[b][b]{$Y_1^{k-1}$}
 \epsfig{file=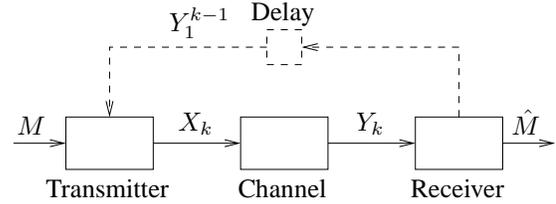, width=0.4\textwidth}
 \caption{The communication system.}
 \label{fig1}
\end{figure}

We consider the communication system depicted in
Figure~\ref{fig1}. The message $M$ to be transmitted over the channel
is assumed to be uniformly distributed over the set
$\set{M}=\{1,\ldots,|\set{M}|\}$ for some positive integer
$|\set{M}|$. The encoder maps the message to the length-$n$ sequence
$X_1,\ldots,X_n$, where $n$ is called the \emph{block-length}. Thus,
in the absence of feedback, the sequence $X_1^n$ is a function of the message
$M$, i.e., $X_1^n=\phi_n(M)$ for some mapping $\phi_n:
\set{M} \to \Reals^n$. Here, $A_m^n$ stands for $A_m,\ldots,A_n$, and
$\Reals$ denotes the set of real numbers.
If there is a feedback link, then $X_k$, $k=1,\ldots,n$, is a function of
the message $M$ and, additionally, of the past channel output symbols
$Y_1^{k-1}$, i.e., $X_k=\varphi_n^{(k)}(M,Y_1^{k-1})$ for some mapping
$\varphi_n^{(k)}: \set{M} \times \Reals^{k-1} \to \Reals$.
The receiver guesses the
transmitted message $M$ based on the $n$ channel output symbols
$Y_1^n$, i.e., $\hat{M}=\psi_n(Y_1^n)$ for some mapping $\psi_n: \Reals^n \to \set{M}$.

Let $\Integers^+$ denote the set of positive integers.
The channel output $Y_k \in \Reals$ at time $k\in\Integers^{+}$
corresponding to the channel inputs $(x_1,\ldots,x_k) \in \Reals^k$ is given
by
\begin{equation}
  Y_k = x_k + \sqrt{\left(\sigma^2+\sum_{\nu=1}^{k-1} \alpha_{k-\nu} x_{\nu}^2\right)} \cdot U_k\label{eq:channel}
\end{equation}
where $\{U_k\}$ are independent and identically distributed (IID),
zero-mean, unit-variance Gaussian random variables drawn independently
of $M$.
The coefficients $\{\alpha_{\nu}\}$
are non-negative and satisfy\footnote{For on-chip communication the coefficients
  characterize the cool-down behavior of the chip, and it thus seems
reasonable to assume that the coefficients are monotonically
non-increasing, i.e., $\alpha_{\nu} \leq \alpha_{\nu'}$ for $\nu \geq
\nu'$. This assumption is, however, not required for the results stated in
this paper.}
\begin{equation}
  \sum_{\nu=1}^{\infty} \alpha_{\nu} \triangleq \alpha < \infty. \label{eq:alpha}
\end{equation}
Note that this channel is not stationary as the variance of the
additive noise depends on the time-index $k$.

We study the above channel under an average-power
constraint on the inputs, i.e.,
\begin{equation}
  \frac{1}{n} \sum_{k=1}^n \E{X_k^2} \leq \const{P}, \label{eq:power}
\end{equation}
and we define the signal-to-noise ratio (SNR) as
\begin{equation}
  \SNR \triangleq \frac{\const{P}}{\sigma^2}. \label{eq:snr}
\end{equation}

\subsection{Capacity per Unit Cost}
\label{sub:unitcost}
Let the \emph{rate} $R$ (in nats per channel use) be defined as
\begin{equation}
  R \triangleq \frac{\log |\set{M}|}{n}
\end{equation}
where $\log(\cdot)$ denotes the natural
logarithm function.
A rate is said to be \emph{achievable} if there exists a sequence
of mappings $\phi_n$ (without feedback) or $\varphi_n^{(1)},\ldots,\varphi_n^{(n)}$ (with feedback)
and $\psi_n$ such that the error probability $\Prs{\hat{M}\neq M}$
tends to zero as $n$ goes to infinity. The \emph{capacity} $C$ is the
supremum of all achievable rates. We denote by $C(\SNR)$ the capacity
under the input constraint \eqref{eq:power} when there is no feedback, and we add the subscript ``FB''
to indicate that there is a feedback link. Clearly,
\begin{equation}
  C(\SNR) \leq C_{\textnormal{FB}}(\SNR) \label{eq:noFBtoFB}
\end{equation}
as we can always ignore the feedback link.

In this paper we study
the \emph{capacities per unit cost} which are defined as \cite{verdu90}
\begin{equation}
\dot{C}(0) \triangleq \sup_{\SNR>0}\frac{C(\SNR)}{\SNR} \quad
\textnormal{and} \quad 
\Cfb \triangleq \sup_{\SNR>0} \frac{C_{\textnormal{FB}}(\SNR)}{\SNR}.
\end{equation}
Note that \eqref{eq:noFBtoFB} implies
\begin{equation}
  \dot{C}(0) \leq \Cfb. \label{eq:FBtonoFB}
\end{equation}

\subsection{The Main Result}
\label{sub:result}
Our main result is stated in the following theorem.

\begin{theorem}
  \label{thm:main}
  Consider the above channel model. Then, irrespective of whether feedback is
  available or not, the corresponding capacity per unit cost is given
  by
  \begin{equation}
    \dot{C}(0) = \Cfb = \frac{1}{2}\left(1+\alpha\right)
  \end{equation}
  where $\alpha$ is defined in \eqref{eq:alpha}.
\end{theorem}

Theorem~\ref{thm:main} is proved in Section~\ref{sec:proof}. In
Section~\ref{sec:highSNR} we briefly discuss the above channel at high
SNR. Specifically, we present a sufficient and a necessary condition
on the coefficients $\{\alpha_{\nu}\}$ for capacity to be bounded
in the SNR.

\section{Proof of Theorem \ref{thm:main}}
\label{sec:proof}
In Section~\ref{sub:upperbound} we derive an upper bound on the feedback
capacity $C_{\textnormal{FB}}(\SNR)$, and in Section~\ref{sub:lowerbound} we
derive a lower bound on the capacity $C(\SNR)$ in the absence of
feedback. These bounds are then used in Section~\ref{sub:asymptotic} to derive an upper bound on $\Cfb$ and a lower
bound on $\dot{C}(0)$, and it is shown that both bounds are
equal to $1/2\cdot (1+\alpha)$. Together with \eqref{eq:FBtonoFB} this proves Theorem~\ref{thm:main}.

\subsection{Upper Bound}
\label{sub:upperbound}
As in \cite[Sec. 8.12]{coverthomas91}, the upper bound on
$C_{\textnormal{FB}}(\SNR)$ is based on Fano's inequality and on an upper bound on $\frac{1}{n}
I(M;Y_1^n)$, which for our channel can be expressed, using the chain
rule for mutual information, as
\begin{IEEEeqnarray}{lCl}
  \IEEEeqnarraymulticol{3}{l}{\frac{1}{n} I(M;Y_1^n)}\nonumber\\
  & = & \frac{1}{n}\sum_{k=1}^n
  \left[h(Y_k|Y_1^{k-1})-h(Y_k|Y_1^{k-1},M)\right]\nonumber\\
  & = & \frac{1}{n}\sum_{k=1}^n
  \left[h(Y_k|Y_1^{k-1})-h(Y_k|Y_1^{k-1},M,X_1^k)\right]\nonumber\\
  & = & \frac{1}{n}\sum_{k=1}^n
  \Bigg[h(Y_k|Y_1^{k-1})-h(U_k) \nonumber\\
  & & {} \qquad \quad \;\; -\frac{1}{2}\E{\log\left(\sigma^2+\sum_{\nu=1}^{k-1}\alpha_{k-\nu}X_{\nu}^2\right)}\Bigg]
  \IEEEeqnarraynumspace \label{eq:upper1}
\end{IEEEeqnarray}
where the second equality follows because $X_1^k$ is a function of
$M$ and $Y_1^{k-1}$; and the last equality follows from the behavior
of
differential entropy under translation and scaling \cite[Thms.~9.6.3
\& 9.6.4]{coverthomas91}, and because $U_k$
is independent of $(Y_1^{k-1},M,X_1^k)$.

Evaluating the differential entropy $h(U_k)$ of a Gaussian random
variable, and using the trivial lower bound $\E{\log\left(\sigma^2+\sum_{\nu=1}^{k-1}\alpha_{k-\nu}X_{\nu}^2\right)}
  \geq \log \sigma^2$,
we obtain the final upper bound
\begin{IEEEeqnarray}{lCl}
  \IEEEeqnarraymulticol{3}{l}{\frac{1}{n} I(M;Y_1^n)}\nonumber\\
  & \leq & \frac{1}{n}\sum_{k=1}^n
  \left[h(Y_k|Y_1^{k-1})-\frac{1}{2}\log(2\pi
      e\sigma^2)\right]\nonumber\\
  & \leq & \frac{1}{2} \frac{1}{n}\sum_{k=1}^n
  \log\left(1+\sum_{\nu=1}^{k}\alpha_{k-\nu} \E{X_{\nu}^2}/\sigma^2\right)\nonumber\\
  & \leq &
  \frac{1}{2}\log\left(1+\frac{1}{n}\sum_{k=1}^n\sum_{\nu=1}^k\alpha_{k-\nu}
  \E{X_{\nu}^2}/\sigma^2\right)\nonumber\\
  & = &
  \frac{1}{2}\log\left(1+\frac{1}{n}\sum_{k=1}^n \E{X_{k}^2}/\sigma^2\sum_{\nu=0}^{n-k}\alpha_{\nu}\right)\nonumber\\
  & \leq &
  \frac{1}{2}\log\left(1+(1+\alpha)\frac{1}{n}\sum_{k=1}^n \E{X_{k}^2}/\sigma^2\right)\nonumber\\
  & \leq & \frac{1}{2}\log\left(1+(1+\alpha)\cdot\SNR\right)\label{eq:upper2}
\end{IEEEeqnarray}
where we define $\alpha_0 \triangleq 1$. Here, the second inequality
follows because conditioning cannot increase entropy and from the entropy maximizing property of Gaussian
random variables \cite[Thm.~9.6.5]{coverthomas91}; the next inequality follows by Jensen's inequality;
the following equality by rewriting the double sum; the subsequent inequality
follows because the coefficients are non-negative which implies that
$\sum_{\nu=0}^{n-k}\alpha_{\nu} \leq \sum_{\nu=0}^{\infty}
\alpha_{\nu}=1+\alpha$; and the last inequality follows from the power
constraint \eqref{eq:power}.

\subsection{Lower Bound}
\label{sub:lowerbound}
As aforementioned, the above channel \eqref{eq:channel} is not
stationary, and one therefore needs to exercise some care in relating
the capacity $C(\SNR)$ in the absence of feedback to the quantity
\begin{equation}
  \varliminf_{n \to \infty} \frac{1}{n} \sup I(X_1^n;Y_1^n) \label{eq:quantity}
\end{equation}
(where the maximization is over all input distributions satisfying the
power constraint \eqref{eq:power}). In fact, it is \emph{prima facie}
not clear whether there is a coding theorem associated with \eqref{eq:quantity}.
We shall sidestep this problem by studying the capacity of a different channel whose time-$k$
channel output $\tilde{Y}_k\in\Reals$ is, conditional on the sequence
$\{X_k\}=\{x_k\}$, given by
\begin{equation}
  \tilde{Y}_{k} = x_k +
  \sqrt{\left(\sigma^2+\sum_{\nu=-\infty}^{k-1}\alpha_{k-\nu}x_{\nu}^2\right)}\cdot
  U_k \label{eq:newchannel}
\end{equation}
where $\{U_k\}$ and $\{\alpha_{\nu}\}$ are defined in
Section~\ref{sub:channelmodel}. This channel has the advantage that it
is stationary \& ergodic in the sense that when $\{X_k\}$ is a stationary \& ergodic process then the
pair $\{(X_k,\tilde{Y}_k)\}$ is jointly stationary \& ergodic.
It follows that if the sequences
$\{X_k\}_{k \leq 0}$ and $\{X_k\}_{k \geq 1}$ are independent of each
other, and if the random variables $X_k$, $k \leq 0$, are bounded, then any rate
that can be achieved over this new channel is also achievable over the
original channel. Indeed, the original channel \eqref{eq:channel} can be
converted into
\eqref{eq:newchannel} by adding
\begin{equation*}
  S_k = \sqrt{\left(\sum_{\nu=-\infty}^0 \alpha_{k-\nu}
    X_{\nu}^2\right)}\cdot U_{-k}
\end{equation*}
to the channel output $Y_k$, and, since the independence of $\{X_k\}_{k \leq 0}$ and
$\{X_k\}_{k \geq 1}$ ensures that the sequence $\{S_k\}$ is independent of
the message $M$, it follows that any rate achievable over
\eqref{eq:newchannel} can be achieved over \eqref{eq:channel} by using
a receiver that generates $\{S_k\}$ and
guesses then $M$ based on $\{Y_k+S_k\}_{k=1}^n$.\footnote{The
  boundedness of the random variables $X_k$, $k \leq 0$, guarantees
  that the quantity $\sum_{\nu=-\infty}^0\alpha_{k-\nu}x_{\nu}^2$ is
  finite for any realization of $\{X_k\}_{k \leq 0}$.}

We consider $\{X_{k}\}$ that are block-wise IID in blocks of $L$ symbols. Thus, denoting
$\vect{X}_{\ell}=\trans{(X_{\ell L +1},\ldots,X_{(\ell+1)L})}$ (where
$\trans{(\cdot)}$ denotes the transpose), $\{\vect{X}_{\ell}\}$ are IID with
$\vect{X}_{\ell}$ taking on the value $\trans{(\xi,0,\ldots,0)}$ with
probability $\delta$ and $\trans{(0,\ldots,0)}$ with
probability $1-\delta$, for some $\xi \in \Reals$. Note that to
  satisfy the average-power constraint \eqref{eq:power} we shall
  choose $\xi$ and $\delta$ so that
\begin{equation}
  \frac{\xi^2}{\sigma^2} \delta = L \cdot \SNR. \label{eq:LBpower}
\end{equation}

Let $\tilde{\vect{Y}}_{\ell}=\trans{(\tilde{Y}_{\ell L
  +1},\ldots,\tilde{Y}_{(\ell+1)L})}$, and let $\lfloor \cdot \rfloor$
  denote the floor function. Noting that the pair $\{(\vect{X}_{\ell},\tilde{\vect{Y}}_{\ell})\}$ is jointly
  stationary \& ergodic, it follows from \cite{verduhan94} that the rate
  \begin{equation}
    R = \lim_{n \to \infty} \frac{1}{n}
    I\left(\vect{X}_0^{\lfloor n/L \rfloor -1};\tilde{\vect{Y}}_0^{\lfloor
    n/L \rfloor-1}\right) \label{eq:codingtheorem}
  \end{equation}
  is achievable over the new channel \eqref{eq:newchannel} and, thus,
  yields a lower bound on the capacity $C(\SNR)$ of the original
  channel \eqref{eq:channel}.
We lower bound
$\frac{1}{n}I(\vect{X}_0^{\lfloor n/L \rfloor -1};\tilde{\vect{Y}}_0^{\lfloor
    n/L \rfloor-1})$ as
\begin{IEEEeqnarray}{lCl}
  \IEEEeqnarraymulticol{3}{l}{\frac{1}{n} I\left(\vect{X}_0^{\lfloor n/L \rfloor -1};\tilde{\vect{Y}}_0^{\lfloor
    n/L \rfloor-1}\right)} \nonumber\\
  & = & \frac{1}{n} \sum_{\ell=0}^{\lfloor n/L \rfloor -1}
  I\left(\left.\vect{X}_{\ell};\tilde{\vect{Y}}_0^{\lfloor n/L \rfloor
    -1}\right|\vect{X}_0^{\ell-1}\right)\nonumber\\
  & \geq & \frac{1}{n} \sum_{\ell=0}^{\lfloor n/L \rfloor -1}
  I\left(\left.\vect{X}_{\ell};\tilde{\vect{Y}}_{\ell}\right|\vect{X}_0^{\ell-1}\right)\nonumber\\
  & \geq & \frac{1}{n} \sum_{\ell=0}^{\lfloor n/L \rfloor -1}\left[
  I\left(\left.\vect{X}_{\ell};\tilde{\vect{Y}}_{\ell}\right|\vect{X}_{-\infty}^{\ell-1}\right)\!-\!I\left(\left.\vect{X}_{-\infty}^{-1};\tilde{\vect{Y}}_{\ell}\right|\vect{X}_0^{\ell}\right)\right] \quad \;\;\label{eq:LB1}
\end{IEEEeqnarray}
where we use the chain rule and that reducing observations cannot increase mutual
information. By using that \eqref{eq:alpha} implies
\begin{equation*}
\lim_{\ell \to \infty} \sum_{\nu=\ell}^{\infty} \alpha_{\nu} = 0
\end{equation*}
it can be shown that the second term in the sum on the right-hand side
(RHS) of \eqref{eq:LB1} vanishes as $\ell$ tends to
infinity. This together with a Ces\'aro type theorem
\cite[Thm.~4.2.3]{coverthomas91} yields
\begin{IEEEeqnarray}{lCl}
  \IEEEeqnarraymulticol{3}{l}{\lim_{n \to \infty} \frac{1}{n} I\left(\vect{X}_0^{\lfloor n/L \rfloor -1};\tilde{\vect{Y}}_0^{\lfloor
    n/L \rfloor-1}\right)}
  \nonumber\\
  & \geq & 
    \frac{1}{L}
    I\left(\left.\vect{X}_{0};\tilde{\vect{Y}}_{0}\right|\vect{X}_{-\infty}^{-1}\right)\nonumber\\   
    & & {} -\lim_{n \to \infty} \frac{1}{L}
  \frac{1}{\lfloor n/L \rfloor} \sum_{\ell=0}^{\lfloor n/L \rfloor -1}I\left(\left.\vect{X}_{-\infty}^{-1};\tilde{\vect{Y}}_{\ell}\right|\vect{X}_0^{\ell}\right)
  \nonumber\\
  & = & \frac{1}{L} I\left(\left.\vect{X}_{0};\tilde{\vect{Y}}_{0}\right|\vect{X}_{-\infty}^{-1}\right)\label{eq:LBcesaro}
\end{IEEEeqnarray}
where the first inequality follows by the stationarity of
$\{(\vect{X}_{\ell},\tilde{\vect{Y}}_{\ell})\}$ which implies that
$I(\vect{X}_{\ell};\tilde{\vect{Y}}_{\ell}|\vect{X}_{-\infty}^{\ell-1})$
does not depend on $\ell$, and by noting that, for a fixed $L$, \mbox{$\lim_{n \to \infty}
\frac{\lfloor n/L \rfloor L}{n}=1$}.

We proceed to analyze
$I(\vect{X}_{0};\tilde{\vect{Y}}_{0}|\vect{X}_{-\infty}^{-1}=\vect{x}_{-\infty}^{-1})$
for a given sequence $\vect{X}_{-\infty}^{-1}=\vect{x}_{-\infty}^{-1}$. Making
use of the canonical decomposition of mutual information (e.g., \cite[eq.~(10)]{verdu90}), we have
\begin{IEEEeqnarray}{lCl}
  \IEEEeqnarraymulticol{3}{l}{I\left(\left.\vect{X}_{0};\tilde{\vect{Y}}_{0}\right|\vect{X}_{-\infty}^{-1}=\vect{x}_{-\infty}^{-1}\right)}
  \nonumber\\
  & = & I\left(\left.X_{1};\tilde{\vect{Y}}_{0}\right|\vect{X}_{-\infty}^{-1}=\vect{x}_{-\infty}^{-1}\right)
  \nonumber\\
  & = & \int D\left( \left. f_{\tilde{\vect{Y}}_{0}|X_{1}=x,\vect{x}_{-\infty}^{-1}} \right\| f_{\tilde{\vect{Y}}_{0}|X_{1}=0,\vect{x}_{-\infty}^{-1}} \right) \d  P_{X_{1}}(x)\nonumber\\
  & & {} -D\left(\left.f_{\tilde{\vect{Y}}_{0}|\vect{x}_{-\infty}^{-1}} \right\|
    f_{\tilde{\vect{Y}}_{0}|X_{1}=0,\vect{x}_{-\infty}^{-1}}\right)\nonumber\\
  & = & \delta D\left(\left. f_{\tilde{\vect{Y}}_{0}|X_{1}=\xi,\vect{x}_{-\infty}^{-1}} \right\| f_{\tilde{\vect{Y}}_{0}|X_{1}=0,\vect{x}_{-\infty}^{-1}}\right) \nonumber\\
  & & {} - D\left(\left.f_{\tilde{\vect{Y}}_{0}|\vect{x}_{-\infty}^{-1}} \right\|
    f_{\tilde{\vect{Y}}_{0}|X_{1}=0,\vect{x}_{-\infty}^{-1}}\right)
  \label{eq:LB2}
\end{IEEEeqnarray}
where the first equality
follows because, for our choice of input distribution, $X_{2}=\ldots=X_{L}=0$ and, hence, $X_{1}$ conveys as
much information about $\tilde{\vect{Y}}_{0}$ as $\vect{X}_{0}$.
Here, $D(\cdot\|\cdot)$ denotes relative entropy, and
$f_{\tilde{\vect{Y}}_{0}|X_{1}=\xi,\vect{x}_{-\infty}^{-1}}$,
$f_{\tilde{\vect{Y}}_{0}|X_{1}=0,\vect{x}_{-\infty}^{-1}}$, and
$f_{\tilde{\vect{Y}}_{0}|\vect{x}_{-\infty}^{-1}}$ denote the densities
of $\tilde{\vect{Y}}_{0}$ conditional on the inputs $\left(X_{1}=\xi,\vect{X}_{-\infty}^{-1}=\vect{x}_{-\infty}^{-1}\right)$, $\left(X_{1}=0,\vect{X}_{-\infty}^{-1}=\vect{x}_{-\infty}^{-1}\right)$, and
$\vect{X}_{-\infty}^{-1}=\vect{x}_{-\infty}^{-1}$, respectively. Thus,
$f_{\tilde{\vect{Y}}_{0}|X_{1}=\xi,\vect{x}_{-\infty}^{-1}}$ is the
density of an
$L$-variate Gaussian random vector of mean $\trans{(\xi,0,\ldots,0)}$
and of diagonal covariance matrix $\mat{K}^{(\xi)}_{\vect{x}_{-\infty}^{-1}}$ with diagonal
entries
\begin{IEEEeqnarray*}{lCl}
  \mat{K}^{(\xi)}_{\vect{x}_{-\infty}^{-1}}(1,1) & = &
  \sigma^2+\sum_{i=-\infty}^{-1}\alpha_{-iL}x_{iL+1}^2 \\
  \mat{K}^{(\xi)}_{\vect{x}_{-\infty}^{-1}}(k,k) & = &
  \sigma^2+\alpha_{k-1}\xi^2+\sum_{i=-\infty}^{-1}\alpha_{-iL+k-1}x_{iL+1}^2,\\
  & & \qquad \qquad \qquad \qquad \qquad \qquad k=2,\ldots,L;
\end{IEEEeqnarray*}
$f_{\tilde{\vect{Y}}_{0}|X_{1}=0,\vect{x}_{-\infty}^{-1}}$ is
the density of an $L$-variate, zero-mean Gaussian random vector of diagonal
covariance matrix $\mat{K}^{(0)}_{\vect{x}_{-\infty}^{-1}}$ with diagonal entries
\begin{equation*}
  \mat{K}^{(0)}_{\vect{x}_{-\infty}^{-1}}(k,k) =
  \sigma^2+\sum_{i=-\infty}^{-1}\alpha_{-iL+k-1}x_{iL+1}^2, \quad k=1,\ldots,L;
\end{equation*}
and $f_{\tilde{\vect{Y}}_{0}|\vect{x}_{-\infty}^{-1}}$ is given by
\begin{equation*}
  f_{\tilde{\vect{Y}}_{0}|\vect{x}_{-\infty}^{-1}} = \delta f_{\tilde{\vect{Y}}_{0}|X_{1}=\xi,\vect{x}_{-\infty}^{-1}}+(1-\delta)f_{\tilde{\vect{Y}}_{0}|X_{1}=0,\vect{x}_{-\infty}^{-1}}.
\end{equation*}

In order to evaluate the first term on the RHS of
\eqref{eq:LB2} we note that the relative entropy of two real, $L$-variate Gaussian
random vectors of the respective means $\bfmu_1$ and $\bfmu_2$ and
of the respective covariance matrices $\mat{K}_1$ and $\mat{K}_2$ is
given by
\begin{IEEEeqnarray}{lCl}
  \IEEEeqnarraymulticol{3}{l}{D\left(\Normal{\bfmu_1}{\mat{K}_1}\|\Normal{\bfmu_2}{\mat{K}_2}\right)}\nonumber\\
  & = & \frac{1}{2} \log\det\mat{K}_2 -
  \frac{1}{2}\log\det\mat{K}_1 +
  \frac{1}{2}\trace{\mat{K}_1\mat{K}_2^{-1}-\mat{I}_L} \nonumber\\
  & & {} +
  \frac{1}{2}\trans{(\bfmu_1-\bfmu_2)}\mat{K}_2^{-1}(\bfmu_1-\bfmu_2) \label{eq:DGaussian}
\end{IEEEeqnarray}
with $\det\mat{A}$ and $\trace{\mat{A}}$ denoting
the determinant and the trace of the matrix $\mat{A}$, respectively,
and where $\mat{I}_L$ denotes the $L \times L$ identity matrix. The
second term on the RHS of \eqref{eq:LB2} is analyzed in the next
subsection. 

Let $\E{D(f_{\tilde{\vect{Y}}_{0}|\vect{X}_{-\infty}^{-1}}
    \|f_{\tilde{\vect{Y}}_{0}|X_{1}=0,\vect{X}_{-\infty}^{-1}})}$ denote the second term on
the RHS of \eqref{eq:LB2} averaged over $\vect{X}_{-\infty}^{-1}$,
i.e.,
\begin{IEEEeqnarray}{r}
  \IEEEeqnarraymulticol{1}{l}{\E{D\left(\left.f_{\tilde{\vect{Y}}_{0}|\vect{X}_{-\infty}^{-1}}
        \right\|f_{\tilde{\vect{Y}}_{0}|X_{1}=0,\vect{X}_{-\infty}^{-1}}\right)}}\nonumber\\
  \qquad \qquad \qquad = \E[\vect{X}_{-\infty}^{-1}]{D\left(\left.f_{\tilde{\vect{Y}}_{0}|\vect{x}_{-\infty}^{-1}}
    \right\|f_{\tilde{\vect{Y}}_{0}|X_{1}=0,\vect{x}_{-\infty}^{-1}}\right)}.\nonumber
\end{IEEEeqnarray}
Then, using \eqref{eq:DGaussian} \& \eqref{eq:LB2} and taking expectations
over $\vect{X}_{-\infty}^{-1}$ we obtain, again defining $\alpha_0
\triangleq 1$,
\begin{IEEEeqnarray}{lCl}
  \IEEEeqnarraymulticol{3}{l}{\frac{1}{L}
    I(\vect{X}_{0};\tilde{\vect{Y}}_{0}\left|\vect{X}_{-\infty}^{-1}\right.)}\nonumber\\
  & = & \frac{\delta}{L}\frac{\xi^2}{\sigma^2}\frac{1}{2}\sum_{k=1}^L
  \E{\frac{\alpha_{k-1}}{1+\sum_{i=-\infty}^{-1}
      \alpha_{-iL+k-1} X_{iL+1}^2/\sigma^2}}\nonumber\\
  & & {} -\frac{\delta}{L}\frac{1}{2}\sum_{k=2}^L
  \E{\log\!\left(1\!+\!\frac{\alpha_{k-1}\xi^2}{\sigma^2+\sum_{i=-\infty}^{-1}\alpha_{-iL+k-1}X_{iL+1}^2}\right)}
  \nonumber\\
  & & {} - \frac{1}{L}\E{D\left(\left.f_{\tilde{\vect{Y}}_{0}|\vect{X}_{-\infty}^{-1}} \right\|
      f_{\tilde{\vect{Y}}_{0}|X_{1}=0,\vect{X}_{-\infty}^{-1}}\right)}\nonumber\\
  & \geq & \frac{\delta}{L}\frac{\xi^2}{\sigma^2}\frac{1}{2}\sum_{k=1}^L
  \frac{\alpha_{k-1}}{1+\sum_{i=-\infty}^{-1}
      \alpha_{-iL+k-1} \E{X_{iL+1}^2}/\sigma^2}\nonumber\\
  & & {} -\frac{\delta}{L}\frac{1}{2}\sum_{k=2}^L
  \log\left(1+\alpha_{k-1} \xi^2/\sigma^2\right)
  \nonumber\\
  & & {} - \frac{1}{L}\E{D\left(\left.f_{\tilde{\vect{Y}}_{0}|\vect{X}_{-\infty}^{-1}} \right\|
      f_{\tilde{\vect{Y}}_{0}|X_{1}=0,\vect{X}_{-\infty}^{-1}}\right)}\nonumber\\
  & \geq & \frac{1}{2} \SNR \sum_{k=1}^L
  \frac{\alpha_{k-1}}{1+ \alpha \cdot L\cdot\SNR}\nonumber\\
  & & {} -\frac{1}{2}\SNR \sum_{k=2}^L
  \frac{\log\left(1+\alpha_{k-1} \xi^2/\sigma^2\right)}{\xi^2/\sigma^2}
  \nonumber\\
  & & {} - \frac{1}{L}\E{D\left(\left.f_{\tilde{\vect{Y}}_{0}|\vect{X}_{-\infty}^{-1}} \right\|
      f_{\tilde{\vect{Y}}_{0}|X_{1}=0,\vect{X}_{-\infty}^{-1}}\right)} \label{eq:LBbeforelimit}
\end{IEEEeqnarray}
where the first inequality follows by the lower bound $\E{1/(1+X)}
\geq 1/(1+\E{X})$ which is a consequence of Jensen's inequality applied to
the convex function $1/(1+x)$, $x>0$, and by the upper bound
\begin{IEEEeqnarray*}{l}
  \E{\log\left(1+\frac{\alpha_{k-1}\xi^2}{\sigma^2+\sum_{i=-\infty}^{-1}\alpha_{-iL+k-1}X_{iL+1}^2}\right)}\\
  \qquad \qquad \qquad \qquad \qquad \qquad \quad \leq
  \log\left(1+\alpha_{k-1}  \xi^2/\sigma^2\right)
\end{IEEEeqnarray*}
for every $k=2,\ldots,L$; and the second inequality
follows by \eqref{eq:LBpower} and by upper bounding
\begin{equation*}
  \sum_{i=-\infty}^{-1} \alpha_{-iL+k-1} \leq \sum_{i=1}^{\infty}
  \alpha_i = \alpha
\end{equation*}
for every $k=1,\ldots,L$.

The final lower bound follows now by \eqref{eq:LBbeforelimit} and \eqref{eq:LBcesaro}
\begin{IEEEeqnarray}{lCl}
  \IEEEeqnarraymulticol{3}{l}{\lim_{n \to
  \infty}\frac{1}{n}I\left(\vect{X}_0^{\lfloor n/L
  \rfloor-1};\tilde{\vect{Y}}_0^{\lfloor n/L \rfloor -1}\right)}\nonumber\\
  & \geq & \frac{1}{2} \SNR \sum_{k=1}^L
  \frac{\alpha_{k-1}}{1+ \alpha \cdot L \cdot\SNR} \nonumber\\
  & & {} - \frac{1}{2}\SNR \sum_{k=2}^L
    \frac{\log\left(1+\alpha_{k-1}  \xi^2/\sigma^2\right)}{\xi^2/\sigma^2}
    \nonumber\\
    & & - \frac{1}{L} \E{D\left(\left.f_{\tilde{\vect{Y}}_{0}|\vect{X}_{-\infty}^{-1}} \right\|
        f_{\tilde{\vect{Y}}_{0}|X_{1}=0,\vect{X}_{-\infty}^{-1}}\right)}. \IEEEeqnarraynumspace \label{eq:LBfinal}
\end{IEEEeqnarray}

\subsection{Asymptotic Analysis}
\label{sub:asymptotic}
We start with analyzing the upper bound \eqref{eq:upper2}. We have
\begin{equation}
  \frac{C_{\textnormal{FB}}(\SNR)}{\SNR} \leq
  \frac{\frac{1}{2}\log(1+(1+\alpha)\cdot \SNR)}{\SNR}\leq \frac{1}{2}(1+\alpha)
\end{equation}
where the second inequality follows by upper bounding $\log(1+x) \leq
x$, $x>0$, and we thus obtain
\begin{equation}
  \Cfb = \sup_{\SNR>0} \frac{C_{\textnormal{FB}}(\SNR)}{\SNR} \leq \frac{1}{2}(1+\alpha).\label{eq:asymU}
\end{equation}

In order to derive a lower bound on $\dot{C}(0)$ we first note that
\begin{equation}
  \dot{C}(0) = \sup_{\SNR >0}\frac{C(\SNR)}{\SNR} \geq \lim_{\SNR
  \downarrow 0} \frac{C(\SNR)}{\SNR} \label{eq:concave}
\end{equation}
and proceed by analyzing the limiting ratio of the lower bound
\eqref{eq:LBfinal} to the SNR as the SNR tends to zero.

To this end, we first shall show that
\begin{equation}
  \lim_{\SNR \downarrow 0} \frac{\E{D\left(\left.f_{\tilde{\vect{Y}}_{0}|\vect{X}_{-\infty}^{-1}} \right\|
        f_{\tilde{\vect{Y}}_{0}|X_{1}=0,\vect{X}_{-\infty}^{-1}}\right)}}{\SNR} = 0. \label{eq:o(snr)}
\end{equation}
It was shown in \cite[p.~1023]{verdu90} that for any pair of densities $f_0$
and $f_1$ satisfying $D(f_1\|f_0) < \infty$
\begin{equation}
  \lim_{\beta \downarrow 0} \frac{D\left(\left.\beta f_1 + (1-\beta) f_0\right\|
      f_0\right)}{\beta} =0.\label{eq:anydensities}
\end{equation}
Thus, for any given $\vect{X}_{-\infty}^{-1}=\vect{x}_{-\infty}^{-1}$,
\eqref{eq:anydensities} together with $\delta = \SNR \cdot L
\cdot \sigma^2/\xi^2$ implies that
\begin{equation}
  \lim_{\SNR \downarrow 0} \frac{D\left(\left.f_{\tilde{\vect{Y}}_{0}|\vect{x}_{-\infty}^{-1}} \right\|
      f_{\tilde{\vect{Y}}_{0}|X_{1}=0,\vect{x}_{-\infty}^{-1}}\right)}{\SNR} = 0. \label{eq:forany}
\end{equation}
In order to show that this also holds when $D(f_{\tilde{\vect{Y}}_{0}|\vect{x}_{-\infty}^{-1}} \|
  f_{\tilde{\vect{Y}}_{0}|X_{1}=0,\vect{x}_{-\infty}^{-1}})$ is averaged over
$\vect{X}_{-\infty}^{-1}$,
we derive in the following the uniform upper bound
\begin{IEEEeqnarray}{lCl}
  \IEEEeqnarraymulticol{3}{l}{\sup_{\vect{x}_{-\infty}^{-1}} D\left(\left.f_{\tilde{\vect{Y}}_{0}|\vect{x}_{-\infty}^{-1}} \right\|
      f_{\tilde{\vect{Y}}_{0}|X_{1}=0,\vect{x}_{-\infty}^{-1}}\right)}\nonumber\\
  \IEEEeqnarraymulticol{3}{r}{\qquad \qquad {} = D\left.\left(\left.f_{\tilde{\vect{Y}}_{0}|\vect{x}_{-\infty}^{-1}} \right\|
      f_{\tilde{\vect{Y}}_{0}|X_{1}=0,\vect{x}_{-\infty}^{-1}}\right)\right|_{\vect{x}_{-\infty}^{-1}=0}. \IEEEeqnarraynumspace} \label{eq:uniformbound}
\end{IEEEeqnarray}
The claim \eqref{eq:o(snr)} follows
then by upper bounding
\begin{IEEEeqnarray}{lCl}
  \IEEEeqnarraymulticol{3}{l}{\E{D\left(\left.f_{\tilde{\vect{Y}}_{0}|\vect{X}_{-\infty}^{-1}} \right\|
        f_{\tilde{\vect{Y}}_{0}|X_{1}=0,\vect{X}_{-\infty}^{-1}}\right)} }\nonumber\\
  \IEEEeqnarraymulticol{3}{r} {\qquad \qquad {} \leq D\left.\left(\left.f_{\tilde{\vect{Y}}_{0}|\vect{x}_{-\infty}^{-1}} \right\|
      f_{\tilde{\vect{Y}}_{0}|X_{1}=0,\vect{x}_{-\infty}^{-1}}\right)\right|_{\vect{x}_{-\infty}^{-1}=0} \IEEEeqnarraynumspace}
\end{IEEEeqnarray}
and by \eqref{eq:forany}.

In order to prove \eqref{eq:uniformbound} we use that any Gaussian random
vector can be expressed as the sum of two independent Gaussian random vectors to
write the channel output $\tilde{\vect{Y}}_{0}$ as
\begin{equation}
  \tilde{\vect{Y}}_{0} = \vect{X}_{0}+\vect{V}+\vect{W}
\end{equation}
where, conditional on $\vect{X}_{-\infty}^{0}=\vect{x}_{-\infty}^{0}$, $\vect{V}$
and $\vect{W}$ are $L$-variate,
zero-mean Gaussian random vectors, drawn independently of each other,
and having
the respective diagonal covariance matrices
$\mat{K}_{\vect{V}|\vect{x}_{0}}$ and
$\mat{K}_{\vect{W}|\vect{x}_{-\infty}^{-1}}$
whose diagonal entries are given by
\begin{IEEEeqnarray*}{lCl}
  \mat{K}_{\vect{V}|\vect{x}_{0}}(1,1) & = & \sigma^2\\
  \mat{K}_{\vect{V}|\vect{x}_{0}}(k,k) & = & \sigma^2 +
  \alpha_{k-1}x_{1}, \qquad k=2,\ldots,L,
\end{IEEEeqnarray*}
and
\begin{equation*}
  \mat{K}_{\vect{W}|\vect{x}_{-\infty}^{-1}}(k,k) = \sum_{i=-\infty}^{-1}
  \alpha_{-iL+k-1}x_{iL+1}^2, \quad k=1,\ldots,L.
\end{equation*}
Thus, $\vect{V}$ is the portion of the noise due to
$\vect{X}_{0}$, and $\vect{W}$ is the portion of the noise due
to $\vect{X}_{-\infty}^{-1}$.
Note that $\vect{X}_{0}+\vect{V}$ and $\vect{W}$ are
independent of each other because $\vect{X}_{0}$ is, by construction, independent
of $\vect{X}_{-\infty}^{-1}$.

The upper bound \eqref{eq:uniformbound} follows now by
\begin{IEEEeqnarray}{lCl}
  \IEEEeqnarraymulticol{3}{l}{D\left(\left.f_{\tilde{\vect{Y}}_{0}|\vect{x}_{-\infty}^{-1}} \right\|
      f_{\tilde{\vect{Y}}_{0}|X_{1}=0,\vect{x}_{-\infty}^{-1}}\right)} \nonumber\\
  & = & D\left(\left.f_{\vect{X}_{0}+\vect{V}+\vect{W}|\vect{x}_{-\infty}^{-1}} \right\|
    f_{\vect{X}_{0}+\vect{V}+\vect{W}|X_{1}=0,\vect{x}_{-\infty}^{-1}}\right)\nonumber\\
  & \leq & D\left(\left.f_{\vect{X}_{0}+\vect{V}} \right\|
    f_{\vect{X}_{0}+\vect{V}|X_{1}=0}\right)\nonumber\\
  & = & D\left.\left(\left.f_{\tilde{\vect{Y}}_{0}|\vect{x}_{-\infty}^{-1}} \right\|
      f_{\tilde{\vect{Y}}_{0}|X_{1}=0,\vect{x}_{-\infty}^{-1}}\right)\right|_{\vect{x}_{-\infty}^{-1}=0}\label{eq:asym1}
\end{IEEEeqnarray}
where
$f_{\vect{X}_{0}+\vect{V}+\vect{W}|\vect{x}_{-\infty}^{-1}}$ and $f_{\vect{X}_{0}+\vect{V}+\vect{W}|X_{1}=0,\vect{x}_{-\infty}^{-1}}$ 
denote the densities of $\vect{X}_{0}+\vect{V}+\vect{W}$
conditional on the inputs $\vect{X}_{-\infty}^{-1}=\vect{x}_{-\infty}^{-1}$
and $(X_{1}=0,\vect{X}_{-\infty}^{-1}=\vect{x}_{-\infty}^{-1})$,
respectively; $f_{\vect{X}_{0}+\vect{V}}$ denotes
the unconditional density of
$\vect{X}_{0}+\vect{V}$; and $f_{\vect{X}_{0}+\vect{V}|X_{1}=0}$ denotes the density of
$\vect{X}_{0}+\vect{V}$ conditional on $X_{1}=0$.
Here, the
inequality follows by the data processing inequality for relative
entropy (see \cite[Sec.~2.9]{coverthomas91}) and by noting that
$\vect{X}_{0}+\vect{V}$ is independent of $\vect{X}_{-\infty}^{-1}$.

Returning to the analysis of \eqref{eq:LBfinal}, we obtain from
\eqref{eq:concave} and \eqref{eq:o(snr)}
\begin{IEEEeqnarray}{lCl}
  \IEEEeqnarraymulticol{3}{l}{\dot{C}(0)}\nonumber\\ 
  & \geq & \lim_{\SNR \downarrow 0} \frac{1}{2} \sum_{k=1}^L
  \frac{\alpha_{k-1}}{1+ \alpha \cdot L \cdot\SNR} - \frac{1}{2} \sum_{k=2}^L
  \frac{\log\left(1+\alpha_{k-1}\frac{\xi^2}{\sigma^2}\right)}{\xi^2/\sigma^2}\nonumber\\
  & = & \frac{1}{2} \sum_{k=1}^{L} \alpha_{k-1} - \frac{1}{2} \sum_{k=2}^L
  \frac{\log\left(1+\alpha_{k-1}
  \xi^2/\sigma^2\right)}{\xi^2/\sigma^2}.
\end{IEEEeqnarray}
By letting first $\xi^2$ go to infinity while holding $L$ fixed, and by letting
then $L$ go to infinity, we obtain the desired lower bound on the
capacity per unit cost
\begin{equation}
  \dot{C}(0) \geq \frac{1}{2} (1+\alpha).\label{eq:asymL}
\end{equation}
Thus, \eqref{eq:asymL}, \eqref{eq:FBtonoFB}, and \eqref{eq:asymU} yield
\begin{equation}
  \frac{1}{2} (1+\alpha) \leq \dot{C}(0) \leq \Cfb \leq \frac{1}{2}(1+\alpha)
\end{equation}
which proves Theorem~\ref{thm:main}.

\section{High SNR Results}
\label{sec:highSNR}
The channel described in Section~\ref{sub:channelmodel} was studied at
high SNR in
\cite{kochlapidothsotiriadis07_2_submitted_to} where it was asked 
whether capacity is bounded or unbounded in
the SNR. It was shown that the answer to this question depends
highly on the decay rate of the coefficients $\{\alpha_{\nu}\}$. We summarize the main result of
\cite{kochlapidothsotiriadis07_2_submitted_to} in the next theorem. For a statement of this
theorem in its full generality and for a proof thereof we refer
to \cite{kochlapidothsotiriadis07_2_submitted_to}.

\begin{theorem}
  \label{thm:highSNR}
  Consider the channel model described in
  Section~\ref{sub:channelmodel}. Then,
  \begin{IEEEeqnarray}{llCl}
    \textnormal{i)} \quad & \varliminf_{\nu \to \infty}
    \frac{\alpha_{\nu+1}}{\alpha_{\nu}} > 0 \quad & \Longrightarrow &
    \quad \sup_{\SNR > 0} C_{\textnormal{FB}}(\SNR) <
    \infty,\label{eq:i} \IEEEeqnarraynumspace\\
    \textnormal{ii)} & \varlimsup_{\nu \to \infty} \frac{\alpha_{\nu+1}}{\alpha_{\nu}}= 0 &
    \Longrightarrow & \quad \sup_{\SNR >0} C(\SNR) = \infty,\label{eq:ii}
  \end{IEEEeqnarray}
  where we define, for any $a>0$, $a/0 \triangleq \infty$ and $0/0 \triangleq 0$.
\end{theorem}

For example, when $\{\alpha_{\nu}\}$ is a geometric sequence, i.e., $\alpha_{\nu}=\rho^{\nu}$ for $0<\rho<1$, then
the capacity is bounded. Note that when neither the
left-hand side (LHS) of \eqref{eq:i} nor the
LHS of \eqref{eq:ii} holds, i.e., when $\varlimsup_{\nu \to \infty}
\alpha_{\nu+1}/\alpha_{\nu}>0$ and $\varliminf_{\nu \to \infty}
\alpha_{\nu+1}/\alpha_{\nu}=0$, then the capacity can be bounded or unbounded.

\section*{Acknowledgment}
Fruitful discussions with Ashish Khisti are gratefully acknowledged.


\end{document}